\documentclass[structabstract]{aa}  

\usepackage{graphicx}
\usepackage{natbib}
\usepackage{txfonts}
\usepackage{url}
\usepackage{multirow}

\begin{document}

\title{Metallicity Gradients in Disks}
\subtitle{Do Galaxies Form Inside-Out?}
\author{K. Pilkington \inst{1,2,3}\fnmsep\thanks{These authors contributed equally to this work.}
       \and C.~G. Few \inst{1,2,\star}
       \and B.~K. Gibson \inst{1,2,3}
       \and F. Calura \inst{1,4}
       \and L. Michel-Dansac \inst{5}
       \and R.~J. Thacker \inst{2}
       \and M. Moll\'{a} \inst{1,6}
       \and F. Matteucci \inst{7}
       \and A. Rahimi \inst{8}
       \and D. Kawata \inst{8}
       \and C. Kobayashi \inst{9}
       \and C.~B. Brook \inst{1,10}
       \and G.~S. Stinson \inst{1,11}
       \and H.~M.~P. Couchman \inst{12}
       \and J. Bailin \inst{13}
       \and J. Wadsley \inst{12}
}

     \institute{Jeremiah Horrocks Insitute, University of Central Lancashire,
       Preston, PR1~2HE, UK\\
       \email{kpilkington@uclan.ac.uk}
       \email{cgfew@uclan.ac.uk}
       \and Department of Astronomy \& Physics, Saint Mary's University, 
            Halifax, Nova~Scotia, B3H~3C3, Canada
       \and Monash Centre for Astrophysics, School of Mathematical Sciences, 
            Monash University, Clayton, Victoria, 3800, Australia
       \and INAF, Osservatorio Astronomico di Bologna, via Ranzani 1,
            Bologna, 40127, Italy     
       \and Centre de Recherche Astrophysique de Lyon, Universit\'{e} de Lyon, 
            Universit\'{e} Lyon 1, Observatoire de Lyon, Ecole Normale 
            Sup\'{e}rieure de Lyon, CNRS, UMR 5574, 9 avenue Charles Andr\'{e},
            Saint-Genis Laval, 69230, France
       \and Departamento de Investigaci\'{o}n B\'{a}sica, CIEMAT, Avda. 
            Complutense 22, Madrid, E28040, Spain
       \and Departimento di Fisica, Sezione di Astronomia, Universit\`{a} di 
            Trieste, via G.B.  Tiepolo 11, 34131, Trieste, Italy
       \and Mullard Space Science Laboratory, University College London, 
            Holmbury St. Mary, Dorking, Surrey, RH5~6NT, UK
       \and Centre for Astrophysics Research, University of Hertfordshire,
            Hatfield, AL10~9AB, UK
       \and Departamento de F\'isica Te\'orica, Universidad Aut\'onoma de
            Madrid, Cantoblanco, Madrid, E28049, Spain
       \and Max-Planck-Institut f\"ur Astronomie, K\"onigstuhl 17,
            Heidelberg, 69117, Germany
       \and Department of Physics and Astronomy, McMaster University, Hamilton,
            Ontario, L8S~4M1, Canada
       \and Astronomy Department, University of Michigan, 500 Church St., Ann 
            Arbor, MI, 48109-1042, USA
}

\abstract
{}
{We examine radial and vertical metallicity gradients using a suite of 
disk galaxy hydrodynamical simulations, supplemented with two classic 
chemical evolution approaches. We determine the rate of change of gradient 
slope and reconcile the differences existing between extant models and 
observations within the canonical ``inside-out'' disk growth paradigm.}
{A suite of 25 cosmological disks is used to examine the evolution of 
metallicity gradients; this consists of 19 galaxies selected from the 
RaDES (Ramses Disk Environment Study) sample (Few et~al., in prep), 
realised with the adaptive mesh refinement code \textsc{ramses}, 
including eight drawn from the `field' and six from `loose group' 
environments. Four disks are selected from the MUGS (McMaster Unbiased 
Galaxy Simulations) sample (Stinson et~al. 2010), generated with the 
smoothed particle hydrodynamics (SPH) code \textsc{gasoline}, 
alongside disks from Rahimi et~al. (2011: \textsc{gcd+}) and
Kobayashi \& Nakasato (2011: \textsc{grape-SPH}).
Two chemical evolution models of 
inside-out disk growth (Chiappini et~al. 2001; Moll\'{a} \& D\'{\i}az 
2005) were employed to contrast the temporal evolution of their radial 
gradients with those of the simulations.}
{We first show that generically flatter gradients are observed at 
redshift zero when comparing older stars with those forming today, 
consistent with expectations of kinematically hot simulations, 
but counter to that observed in the Milky Way. 
The vertical abundance gradients at 
$\sim$1$-$3 disk scalelengths are comparable to those observed in the 
thick disk of the Milky Way, but significantly shallower than those seen 
in the thin disk.  Most importantly, we find that systematic differences 
exist between the predicted evolution of radial abundance gradients in 
the RaDES and chemical evolution models, compared with the MUGS sample; 
specifically, the MUGS simulations are systematically steeper at 
high-redshift, and present much more rapid evolution in their 
gradients.}
{We find that the majority of the models predict radial gradients today 
which are consistent with those observed in late-type disks, but they 
evolve to this self-similarity in different fashions, despite each 
adhering to classical `inside-out' growth.  We find that radial 
dependence of the efficiency with which stars form as a function of time 
drives the differences seen in the gradients; systematic differences in 
the sub-grid physics between the various codes are responsible for 
setting these gradients. Recent, albeit limited, data at redshift 
$z$$\sim$1.5 are consistent with the steeper gradients seen in our 
SPH sample, suggesting a modest revision of the classical chemical 
evolution models may be required.}


\maketitle

\section{Introduction}

The recognition that metals are not distributed homogeneously throughout 
the disk of the Milky Way \citep{Shav83} has proven to be fundamental in 
our efforts to understand the role of interactions, mergers, accretion, 
migration, and gas flows, in shaping the formation and evolution of 
galaxies.  A rich literature now exists which confirms these radial 
abundance trends in both spirals \citep[e.g.][]{Simp95,Affler97,MHB99, 
Carr08,Kew10,San11} and ellipticals 
\citep[e.g.][]{Korm89,FI90,Pel90}. Vertical trends have been studied 
somewhat less frequently \citep[e.g.][]{Mar05,Mar06,Sou08,Nav11}, but 
provide unique insights into the discrete nature (or lack thereof) of 
the thin disk -- thick disk interface (and associated kinematical 
heating processes).

Observations of nearby spiral galaxies show that the inner disks have 
higher metallicities than their associated outer disk regions; at the 
present day, typical gradients of $\sim$$-$0.05~dex/kpc are encountered. 
These somewhat shallow gradients have provided \it critical \rm 
constraints on models of galaxy formation and evolution, and are 
fundamental to the predictions of the classical ``inside-out'' paradigm 
for disk growth. Predictions have been made of the time evolution of 
metallicity gradients in chemical evolution models 
\citep[e.g.][]{Moll97,Fu09} and observationally from plantetary nebulae 
\citep[e.g.][]{Mac03}, although until recently, we have had essentially 
no \it direct \rm observational constraints on what the magnitude of the 
time evolution of the gradients should be. This has changed with the 
work of \citet{Cres10}, \citet{Jones10}, \citet{Queyrel11}, 
and \citet{Yuan11}, who have, 
for the first time, extended radial abundance gradient work to high 
redshifts.  \citet{Yuan11} show that for at least one 
``Grand Design'' disk at 
redshift $z$$\sim$1.5, the metallicity gradient is significantly steeper 
($-$0.16~dex/kpc) than the typical gradient encountered 
today.\footnote{ At even 
higher redshifts ($z$$\sim$3.3), 
\citet{Cres10} and Troncoso et~al. (2012, in prep), as part of the AMAZE/LSD
surveys, suggest that {\it both \rm} inverted gradients (higher 
abundances in the outskirts, relative to the inner disk) {\it and\rm}
standard declining gradients are seen.  From the latter surveys,
inverted gradients
(ranging from $+$0.0 to $+$0.1~dex/kpc) appear associated with
very massive stellar disks at these high-redshifts
(M$_\ast$$>$3$\times$10$^9$~M$_\odot$), while declining gradients
(ranging from $-$0.0 to $-$0.2~dex/kpc) appear associated with
lower mass stellar disks (M$_\ast$$<$3$\times$10$^9$~M$_\odot$).
\citet{Cres10} suggest that the inverted gradients are
due perhaps to recent infall of pristine material into the inner disk. 
These Lyman Break Galaxies, with their $\sim$1$-$2 orders of magnitude
greater star formation rates (relative to the typical Milky Way
progenitor at that redshift), are more likely associated with 
massive spheroids in clusters/groups today \citep[e.g.][]{Nagamine2002},
as opposed to the Milky Way, and
so are not directly comparable with the simulations described here.}
Constraining the metallicity gradients of galaxies beyond the local 
Universe remains a challenge for the future.

Using SPH simulations of disk galaxy mergers, \citet{Rupke10} show 
strong correlations of metallicity with environment and merger history, 
focussing on the effects of gas inflows and star formation rate. 
Observations by \citet{Coop08} show that higher metallicity galaxies are 
more abundant in group enviroments and \citet{Kew06} showed that 
interacting pairs of galaxies have systematically lower metallicities 
($\sim$0.2~dex lower) than field galaxies or more loosely associated 
pairs. Radial gradients have been shown to flatten for galaxies that 
have experienced recent mergers \citep{Kew10}; these also result in 
higher velocity dispersions and redistribution of the cold gas. In 
agreement with this, \citet{MD08} studied the mass-metallicity relation 
for merging galaxies and concluded that the infall of metal poor gas 
during merger events lowers the gas phase metallicity. However, the 
timescale over which redistributed gas develops into a gradient like 
those we see in spiral galaxies today is unknown.

There have been several studies of chemistry within cosmological 
hydrodynamical simulations \citep[e.g.][]{Rait96, 
Kaw03,ONB08,Scan08a,Zol10,Rah10,Wier11,Kob11}, each modelling certain 
observational properties with varying degrees of success. Some studies 
have examined the radial and/or vertical gradients using hydrodynamical 
codes \citep[e.g.][]{Rupke10,Rah11}, but the numerical study of radial 
gradients has predominantly been in the context of classical galactic 
chemical evolution codes \citep[e.g.][]{Prant00,Chiap01, Moll05}. In 
this paper, we use 25 simulations realised with three different 
cosmological hydrodynamical codes: \textsc{gasoline} \citep{Wad04} and 
\textsc{gcd+} \citep{Kaw03}, both gravitational N-Body $+$ Smoothed 
Particle Hydrodynamic (SPH) \citep{Mon92} codes, and \textsc{ramses} 
\citep{Tey02}, an Adaptive Mesh Refinment (AMR) code. Alongside these, 
we use the results from the chemical evolution models of \citet{Chiap01} 
and \citet{Moll05}.

Our work aims to fill an important gap in the field, by complementing 
orbital parameter studies \citep[e.g.][]{Rupke10,Perez11}, systematic 
sub-grid physics parameter studies \citep[e.g.][]{Wier11}, and detailed 
dissections of individual systems \citep[e.g.][]{Rah11,Zol10,Kob11}, 
with a statistical sample of Milky Way-like analogs. 
Our approach is 
different, but complementary, to the careful and compelling parameter 
study of \citet{Wier11}; their, the goal was to vary the input physics 
and examine the outcome, regardless of whether or not the simulated 
end-products might be classified still as Milky Way-like. Instead, we 
have sampled a range of codes, sub-grid physics, and initial conditions, 
each of which has been `calibrated', in some sense, by their respective 
authors, to resemble a classical Milky Way-like system. With that 
calibrated sample, our unique contribution is to examine the `path' by 
which the gradients evolve, search for both random and systematic 
trends/differences between the samples, and compare with new empirical 
data at high-redshift.\footnote{In spirit, this is exactly the approach 
taken in the seminal Galactic Chemical Evolution Comparison Project 
\citep{Tosi1996}, which examined the time evolution of classic chemical 
evolution models {\it calibrated \rm} to the solar neighbourhood, in order 
to see where they differed `away' from this calibrated boundary 
condtion.} {\it This is the first time such a comparison of the temporal 
evolution of metallicity gradients has been undertaken with a 
statistical sample of simulated disk galaxies.  \rm}

The outline of the paper is as follows. The main differences between the 
codes are described in \S\ref{sims}, where we concentrate primarily upon 
the relevant mechanisms associated with the treatment of star formation 
and feedback (both energetic and chemical). The metallicity gradients 
inferred today for stellar populations of different ages are presented 
in \S\ref{presGrad}. This is expanded upon in \S\ref{evo} where the 
radial metallicity gradients of the young stellar population as a 
function of redshift are considered. Finally, we summarise our findings 
in \S\ref{summ}.

\section{Simulations}
\label{sims}

The simulations used in this paper are fully described in Stinson et~al. 
(2010: MUGS), Rahimi et~al. (2011: Gal1), Kobayashi \& Nakasato
(2011: KN11) and Few et~al. (in prep: 
RaDES);  the main characteristics of the simulations and their parent 
codes are described here and itemised in Table~\ref{tabvals}. The 
chemical evolution models are fully described in \citet{Chiap01} and 
\citet{Moll05}, but again we describe the main aspects in the following 
section.

\begin{table*}
\centering
\begin{tabular}{ l  l  r  r  c  c  c  c }
  \hline
 Suite & Galaxy & M$_{\rm Tot}$        & M$_{\rm *,disk}$    & r$_{\rm disk}$ & Environment & d[$Z_{\rm *,all}$]/d$h$ & d[$Z_{\rm *,young}$]/d$R$ \\
       & Galaxy & (10$^{11}$M$_\odot$) & (10$^{10}$M$_\odot$) & (kpc)         &             & (dex/kpc)               & (dex/kpc)                 \\
  \hline
  \multirow{4}{*}{MUGS}   & g15784     & 14.0 &  5.9  & 3.2 & Field       & $-$0.06 & $-$0.04 \\
                          & g422       &  9.1 &  2.0  & 2.8 & Field       & $-$0.06 & $-$0.08 \\
                          & g1536      &  7.0 &  3.3  & 2.5 & Field       & $-$0.07 & $-$0.05 \\
                          & g24334     &  7.7 &  2.7  & 1.0 & Field       & $-$0.03 & $-$0.19 \\
  \hline
  \multirow{1}{*}{GCD$+$} & Gal1       &  8.8 &  4.1  & 2.7 & Field       & $-$0.04 & $-$0.01 \\
  \hline
  \multirow{1}{*}{Grape-SPH} & KN11    & 11.0 &  2.0  & 4.7 & Field       & $-$0.03 & $-$0.04 \\
  \hline
  \multirow{19}{*}{RaDES} & Castor     & 10.5 &  7.2  & 4.0 & Loose Group & $-$0.17 & $-$0.03 \\
                          & Pollux     &  4.2 &  3.4  & 3.0 & Loose Group & $-$0.06 & $-$0.05 \\
                          & Tyndareus  &  3.3 &  1.3  & 1.3 & Loose Group & $-$0.02 & $-$0.05 \\
                          & Zeus       &  2.3 &  1.0  & 1.7 & Loose Group & $-$0.07 & $-$0.04 \\
                          & Apollo     &  8.9 &  6.3  & 3.0 & Loose Group & $-$0.04 & $-$0.06 \\
                          & Artemis    &  7.5 &  3.2  & 1.9 & Loose Group & $-$0.08 & $-$0.05 \\
                          & Daphne     &  3.1 &  2.1  & 2.7 & Loose Group & $-$0.03 & $-$0.06 \\
                          & Leto       &  2.5 &  1.2  & 1.8 & Loose Group & $-$0.04 & $-$0.05 \\
                          & Luke       & 11.3 &  6.6  & 5.4 & Loose Group & $-$0.01 & $-$0.03 \\
                          & Leia       &  3.9 &  3.0  & 4.1 & Loose Group & $-$0.05 & $-$0.02 \\
                          & Tethys     &  7.2 &  5.1  & 2.8 & Field       & $-$0.08 & $-$0.05 \\
                          & Krios      &  5.7 &  4.0  & 2.5 & Field       & $-$0.10 & $-$0.05 \\
                          & Atlas      &  6.5 &  4.4  & 2.8 & Field       & $-$0.06 & $-$0.04 \\
                          & Hyperion   & 10.0 &  7.7  & 3.6 & Field       & $-$0.07 & $-$0.04 \\
                          & Eos        &  4.6 &  2.5  & 2.0 & Field       & $-$0.19 & $-$0.07 \\
                          & Helios     & 10.5 &  6.6  & 1.6 & Field       & $-$0.11 & $-$0.04 \\
                          & Selene     &  6.1 &  5.2  & 3.5 & Field       & $-$0.05 & $-$0.06 \\
                          & Oceanus    & 11.0 & 10.0  & 6.6 & Field       & $-$0.03 & $-$0.03 \\
                          & Ben        &  7.7 &  4.2  & 3.9 & Field       & $-$0.04 & $-$0.03 \\
  \hline
\end{tabular}
\caption{Basic present-day ($z$=0) characteristics of the 25 simulated 
disks. Column (1): simulation suite to which the the code used to 
simulate the galaxy (Column (2)) belongs; Column (3): total (dynamical) 
mass within the virial radius; Column (4): mass of the stellar disk, 
after application of the kinematic and spatial cuts described in \S~3; 
Column (5): exponential scalelength of the stellar disk; Column (6): 
local environment of the galaxy; Column (7): mass-weighted vertical 
stellar abundance gradient, averaged over the radial range 1.5$<$$r_{\rm 
disk}$2.5; Column (8): mass-weighted radial {\it young \rm} (stars born 
within the past 100~Myrs) stellar abundance gradient, after application 
of the kinematic and spatial cuts described in \S~3.}
\label{tabvals}
\end{table*}

\subsection{MUGS}

The MUGS galaxies were run using the gravitational N-body $+$ SPH code 
\textsc{gasoline} which was introduced and described in \citet{Wad04}. 
Below, we emphasise the the main points concerning the star formation 
and feedback sub-grid physics used to generate this suite of 
simulations, but first remind the reader of the background framework 
in which they were evolved, in addition to their basic characteristics. 

The MUGS sample \citep{Stinson10} consists of 16 galaxies randomly drawn 
from a cosmological volume 50$h^{-1}$~Mpc on a side, evolved in a 
 Wilkinson Microwave Anisotropy Probe Three (WMAP3)  
$\Lambda$CDM cosmology with $H_0=73$~km~s$^{-1}$~Mpc$^{-1}$, 
$\Omega_m=0.24$, $\Omega_\Lambda=0.76$, $\Omega_b=0.04$, and 
$\sigma_8=0.76$. Each galaxy is resimulated at high resolution by 
using the volume renormalization technique \citep{Kly01}, with a 
gravitational softening length of 310~pc. The galaxies range in mass 
from 5$\times$10$^{11}$~M$_{\odot}$ to 2$\times$10$^{12}$~M$_{\odot}$.  
The four galaxies with the most prominent disks\footnote{ By
`prominent', we mean the inclusion of those for which there was
unequivocal identification of the disk (from angular momentum 
arguments constructed from the gas and young star distributions, 
as discussed in \S3.1.  In a secondary sense, this eliminated
extreme values of bulge-to-total, but formally, we only
included those disks for which alignment based upon the gas/young stars
was obvious.} were selected: 
g422\footnote{g422 was not described in the original MUGS paper 
\citep{Stinson10}; it was produced identically to the MUGS suite and 
will be described fully in an upcoming paper.}, g1536, g24334, and 
g15784, the latter of which is the closest to a Milky Way analog 
in the sample.

Star formation and supernovae feedback uses the blastwave model 
\citep{Stinson06} whereby gas particles can form stars when they are 
sufficiently dense ($>$1~cm$^{-3}$) and cool ($<$15000~K). Gas particles 
which satisfy these criteria can form stars according to the equation 
$\frac{dM_{\star}}{dt}$$=$c$^{\star}$$\frac{M_{gas}}{t_{dyn}}$, where 
c$^{\star}$ is the star formation efficiency and is fixed to be 0.05. 
$M_{gas}$ is the mass of the gas particle forming the star particle of 
mass $M_{\star}$ and $ t_{dyn}$ is the dynamical time of the gas. 
Heating from a uniform ultraviolet ionising background radiation field 
\citep{Haardt96} is employed, and cooling is derived from the 
contributions of both primordial gas and metals; the metal cooling grid 
is derived using \textsc{CLOUDY} (v.07.02: \citet{Ferland98}), under 
the assumption of ionisation of equilibrium, as detailed by 
\citet{Shen2010}.

The chemical evolution model used in \textsc{gasoline} is fully 
described in \citet{Rait96b}; here, we only discuss the main points. All 
stars with masses above 8~M$_{\odot}$ explode as Type~II supernova 
(SNeII). An efficiency factor couples 40\% of a given supernova's energy 
(10$^{51}$~erg) to the surrounding interstellar medium (ISM). The metals 
that are tracked in this version of \textsc{gasoline} (O and Fe) all 
come from supernovae and
are allowed to diffuse between neighbouring SPH particles, after 
\citet{Shen2010}. The Type~Ia supernovae
(SNeIa) eject iron and oxygen; 
for every SNIa, 0.76~M$_{\odot}$ of `metals' is ejected, divided between 
iron (0.63~M$_{\odot}$) and oxygen (0.13~M$_{\odot}$). 
Our binary model for      
Type~Ia supernovae is based upon the single-degenerate                         
progenitor formalism of \citet{Greg83}, with secondaries spanning in 
mass from 1.5 to 8.0~M$_\odot$.\footnote{ We have excluded 
secondaries in the 0.8 - 1.5~M$_\odot$ range; doing so, regardless of IMF,
only impacts on the SNeIa rate at the $\sim$20\% level.\rm}
Enrichment from 
SNeII is based upon power law fits in stellar mass to the 
nucleosynthesis yield tables of \citet{WooWe95}, convolved with a Kroupa 
\citep{Krou93} initial mass function (IMF), in order to determine the 
mass fraction of metals ejected. The total metallicity in this version 
of the code is tracked by assuming Z$\equiv$O+Fe.\footnote{By assuming 
Z=O+Fe, we admittedly underestimate the global metal production rate by 
nearly a factor of two; our next generation runs with \textsc{gasoline} 
employ a more detailed chemical evolution model, incorporating the
nucleosynthetic byproducts of asymptotic giant branch evolution and
thereby ameliorating this effect.} 
For these runs, only the Z=Z$_\odot$ yields were used, and 
long-lived SNeIa progenitors (those with secondaries with mass 
$m$$<$1.5~M$_\odot$) were neglected.

\subsection{Gal1}

Gal1 is a higher-resolution re-simulation of galaxy D1 from 
\citet{Kaw04} using the SPH code \textsc{gcd+} \citep{Kaw03}; while its 
characteristics have been discussed previously by \citet{Bailin05}, 
\citet{Rah10}, and \citet{Rah11}, an overview is provided here for 
completeness.  Employing a comparable volume renormalisation / 
`zoom-style' technique to that described in \S~2.1 (with a gravitational 
softening of 570~pc in the highest resolution region), Gal1 was realised 
within a $\Lambda$CDM cosmological framework with 
$H_0=70$~km~s$^{-1}$~Mpc$^{-1}$, $\Omega_m=0.3$, $\Omega_\Lambda=0.7$, 
$\Omega_b=0.04$, and $\sigma_8=0.9$, resulting in a Milky Way analog of 
virial mass 8.8$\times$10$^{11}$~M$_{\odot}$.  The effect of the 
ultraviolet background radiation field was neglected, while 
metal-dependent radiative cooling (adopted from \textsc{MAPPINGS-III} 
\citep{SuthDop93}) was included.

The star formation prescription employed requires (i) the gas density to 
be above a threshold of 0.1~cm$^{-3}$, (ii) a convergent gas flow to 
exist, and (iii) the gas to be locally Jeans unstable. A standard 
Salpeter initial mass function (IMF) was assumed, along with pure
thermal feedback from both SNeII and SNeIa (10$^{50}$~erg/SN) being 
coupled to the surrounding SPH particles.

The chemical evolution implementation within \textsc{GCD+} takes into 
account the metal-dependent nucleosynthetic byproducts of SNeII 
\citep{WooWe95}, SNeIa \citep{Iwam99}, and low- and intermediate-mass 
AGB stars \citep{Hoek97}.  Relaxing the instantaneous recycling 
approximation, \textsc{GCD+} tracks the temporal evolution of the nine 
dominant isotopes of H, He, C, N, O, Ne, Mg, Si, and Fe. The SNeIa 
progenitor formalism of \citet{Kob00} is adopted.

\subsection{KN11}

KN11 corresponds to the so-called `Wider Region' model described
by \citet{Kob11}, realized used a hybrid \textsc{grape-SPH} code.
This model was drawn from the 5 Milky Way-analogs which
eventuated from a larger suite of 150 
semi-cosmological\footnote{ By `semi-cosmological', we mean that the
simulated field was not large enough to sample the longest waves (and,
as such, underestimate the degree of gravitational tidal torque 
which would otherwise be present in a fully cosmological framework),
and so the initial system is provided with an initial angular momentum
via the application of rigid rotation with a constant spin
parameter $\lambda$=1.} simulations.
The cosmological parameters employed match those of \S2.2, 
and led to a Milky Way analog of mass
1.1$\times$10$^{12}$~M$_{\odot}$.  The effect of the 
ultraviolet background radiation field was included, as was
metal-dependent radiative cooling (adopted from \textsc{MAPPINGS-III} 
\citep{SuthDop93}).

The star formation prescription employed requires (i) the gas density to 
be cooling, (ii) a convergent gas flow to 
exist, and (iii) the gas to be locally Jeans unstable. The star 
formation timescale is chosen to be proportional to the
dynamical timescale ($t_{\rm sf}$$\equiv$$t_{\rm dyn}/c$), where 
the star formation efficiency is chosen to be $c$=0.1.
A standard 
Salpeter initial mass function (IMF) was assumed (with lower and upper
mass limits of 0.07 and 120~M$_\odot$, respectively), along with pure
thermal feedback from both SNeII\footnote{ 50\% of the massive
stars are assumed to end their lives as SNeII, while the remaining
50\% are assumed to end their lives as 10$\times$ more energetic
hypernovae.} and SNeIa ($\sim$10$^{51}$~erg/SN) being 
distributed to the surrounding SPH particles within 1~kpc (weighted
by the SPH kernel).

The chemical evolution implementation within \textsc{grape-SPH} takes into 
account the metal-dependent nucleosynthetic byproducts of SNeII 
\citep{Kob06}, SNeIa \citep{Nomoto97}, and low- and intermediate-mass 
AGB stars \citep{Karakas10}.

\subsection{RaDES}

The third galaxy sample (RaDES: Ramses Disk Environment Study) was 
simulated using the adaptive mesh refinement (AMR) code \textsc{ramses} 
\citep{Tey02}. The motivation behind these simulations was to determine 
the systematic differences between simulated galaxies with neighbouring 
dark matter haloes similar to the Local Group and those in the field. 
The \textsc{ramses} simulations include gravity, radiative cooling, and 
heating from a uniform ionising UV background radiation 
\citep{Haardt96}. Hydrodynamic behaviour of the gas phase and 
gravitational potential is calculated on a spatially adaptive grid. A 
full description of the star formation model used in \textsc{ramses} is 
given by \citet{Dubois08}; here we give just a brief account of its 
implementation.

Gas cells with density greater than a given threshold allow stars to 
form at a rate proportional to the density, $\dot{\rho} = -\rho / 
t_{\star}$, where $t_\star$ is the star formation timescale, which 
itself is proportional to the dynamical time 
($t_0$($\rho/\rho_0)^{-1/2}$), as first described by \citet{Ras06}. 
After \citet{Dubois08}, we use a threshold of $\rho_0$ = 0.1~cm$^{-3}$ 
and t$_0 = $ 8~Gyr. In combination, these choices correspond to an 
adopted star formation efficiency of 2\%. Feedback from 
SNeII\footnote{SNIa are not accounted for in RaDES, although we have 
recently completed a chemical evolution upgrade to \textsc{RAMSES} which 
parallels that implemented within \textsc{GCD+} (\S~2.2); this will 
described in a future work.} occurs instantaneously and the mass carried 
away is parameterised as ($\eta_{SN} + \eta_W$), where $\eta_{SN}$ is 
the fraction of a stellar particle's mass that is ejected by SNeII and 
$\eta_W$ is the fraction that is swept up in the SNII wind. In the RaDES 
simulations, $\eta_{SN}$ = 0.1 and $\eta_W$ = 0, which for these runs, 
led to less strongly peaked rotation curves. Energy is injected into the 
gas phase in the form of thermal and kinetic energy, distributed across 
a superbubble of radius r$_{\rm bubble}$ according to a Sedov blastwave 
formalism.The metallicity of SN ejecta is determined by converting a 
fixed fraction, $f_{Z}$, of the non-metal content of new stars into 
metals; all galaxies in the RaDES sample used $f_{Z}$=0.1.

RaDES is comprised of two subsamples allowing for a statistical 
intercomparison of field galaxies and those in environments similar to 
those of loose groups; the full details will be presented in Few et~al. 
(in prep).  These simulations take place in either 20$h$$^{-1}$~Mpc 
(grid resolution of 440~pc) or 24$h$$^{-1}$~Mpc (grid resolution of 
520~pc) volumes with 512$^3$ dark matter particles in the central 
region. The cosmology of these boxes is H$_0$=70~km~s$^{-1}$Mpc$^{-1}$, 
$\Omega_\Lambda$=0.72, $\Omega_m$=0.28, $\Omega_b$=0.045, and 
$\sigma_8$=0.8. 

The sample employed here consists of nine isolated (field) galaxies and 
ten situated within loose groups.  The latter are defined as being those 
for which two $L_\ast$ halos of comparable mass reside within 1.5~Mpc of 
one another, and neither are located within 5~Mpc of a halo with mass in 
excess of 5$\times$10$^{12}$~M$_\odot$.  The latter criterion avoids the 
proximity to rich clusters.  In a statistical sense, these `loose 
groups' can be thought of as Local Group analogs, at least in terms of 
dynamical mass, proximity to companion galaxies, and the avoidance of 
rich clusters. The field sample contains those halos that are even more 
isolated from neighbouring massive halos : specifically, no M$_{\rm 
halo} >$ 3$\times$10$^{11}$~M$_{\odot}$ within 3~Mpc). The virial mass 
range of the RaDES sample spans 2.5$\times$10$^{11}$ to 
1.6$\times$10$^{12}$~M$_{\odot}$.

\subsection{Chemical Evolution Models}

In this work, we compare our results from the hydrodynamical simulations 
described in \S~2.1--2.3 to two chemical evolution models both designed 
to reproduce the main features of our Galaxy. The models are described 
by \citet{Chiap01} and \citet{Moll05}, and we refer the reader to these 
papers for full details.

In the model by \citet{Chiap01}, the Milky Way forms by means of two 
main infall episodes, both represented by exponential infall rates. The 
first infall episode, characterised by the rate 
$\dot{\sigma_H}$$\propto$$A$~$e$$^{-t/\tau_{inf,H}}$, is associated with 
the formation of the halo and thick disk, with an $e$-folding timescale 
($\tau_{inf,H}$) of $\sim$1~Gyr. The constant $A$ is determined by 
requiring that the present-day mass surface density of the halo is 
reproduced.

The second infall phase is represented as 
$\dot{\sigma_D}$$\propto$$B(R)$~$e^{-t/\tau_{inf,D}}$, and is associated 
with the formation of the thin disk.  The thin disk is represented by 
independent annuli, each 2~kpc wide, with no exchange of matter between 
them (i.e., no radial gas flows).  The $e$-folding timescale 
($\tau_{inf,D}$) of the second infall is assumed to be a linear function 
with increasing galactocentric radius (i.e., $\tau_{inf,D}(R) \propto 
R$) - enforcing the so-called ``inside-out'' paradigm for disk growth, 
with the gas accumulating faster in the inner regions of the disk, 
relative to the outer disk. The timescales here vary from $\sim$2~Gyr in 
the inner disk, to $\sim$7~Gyr in the solar neighbourhood, and up to 
$\sim$20~Gyr in the outermost parts of the disk. The constant $B(R)$ is 
fixed in order to reproduce the present-day total surface mass density 
(stars + gas) in the solar neighbourhood. The star formation rate 
$\dot{\sigma_*}$ is expressed by the common Schmidt-Kennicutt law, 
$\dot{\sigma_*} \propto \nu \sigma_{gas}^k(R,t)$, where 
$\sigma_{gas}(R,t)$ represents the gas density at the radius $R$ and at 
the time $t$, and $k=1.5$. The star formation efficiency $\nu$ is set to 
1~Gyr$^{-1}$, and becomes zero when the gas surface density drops below 
a certain critical threshold, adopted here to be 
$\sigma_{th}$=7~M$_{\odot}~pc^{-2}$.  The nucleosynthesis prescriptions 
for AGB stars and SNeIa+SNeII are drawn from the same sources listed in 
\S~2.2.

The chemical evolution model of \citet{Moll05} differs from that of 
\citet{Chiap01} in several aspects, in that it is multiphase, treating 
the ISM as a mixture of hot diffuse gas and cold molecular clouds. Each 
galaxy is assumed to be a two-zone system, comprised by a halo formed in 
an early gas-rich phase and a disk. The gas of the disk is acquired from 
the halo through an imposed infall prescription characterised by the 
inverse of the collapse time, which itself depends upon the total mass 
of the galaxy. The mass profile is imposed to adhere to the 
\citet{Persic96} universal rotation curve. Similar to \citet{Chiap01}, 
each galaxy is divided into concentric cylindrical zones 1~kpc wide. The 
collapse timescale depends on radius via an exponential function 
$\tau(R) \propto e^R$, rather than the linear dependence upon $R$ 
employed by \citet{Chiap01}. Another important difference concerns the 
treatment of star formation: in the \citet{Moll05} model, stars form in 
two stages: first, molecular clouds condense with some efficiency out of 
the diffuse gas reservoir, and second, stars form with a second 
efficiency factor based upon cloud-cloud collision timescales.  In 
spirit, this mimics the effect of the threshold effect in the 
\citet{Chiap01} model: specifically, stars may form only in dense 
regions. The relation between the star formation rate and the gas 
density can be approximated by a power law with $n>1$, again, in 
qualitative agreement with the law employed by \citet{Chiap01}. In the 
halo, star formation follows a common Schmidt-Kennicutt law with 
exponent $n =1.5$. Extensive testing and tuning of the main parameters 
resulted in a grid of 440 models spanning 44 different masses (from 
dwarfs to giants, with 10 different star formation efficiencies per mass 
model).  The chemical prescriptions for SNeIa and SNeII are again 
similar to those listed in \S~2.2.

\section{Present-Day Gradients}
\label{presGrad}
\subsection{Radial Gradients}
\label{RadGrad}

In this section, the present-day radial abundance gradients of the MUGS 
and RaDES simulations are presented. We focus here on one MUGS (g15784) 
and one RaDES galaxy (Apollo), which have been chosen as fiducial 
representatives of these two suites of simulations. Observational 
constraints on the abundance gradient of $z$=0 late-type galaxies may be 
found in, for example, \cite{Zari94} who measured a mean gradient of 
$-0.058$~dex/kpc for local spiral galaxies and \cite{VanZee98}, who 
found a comparable mean gradient from their sample ($-$0.053~dex/kpc). 
In \cite{Kew10} close galaxy pairs were found to have systematically 
shallower gradients (typically, $-$0.021~dex/kpc).  In each of these 
cases, the gradients are inferred from gas-phase nebular emission, which 
provides a ``snapshot'' of the present-day gradient, similar to that 
inferred from, for example, B-stars (i.e., stars with ages 
$<$100~Myrs).\footnote{Loose group galaxies in the RaDES suite 
exhibit the same qualitative flattening of metallicity gradients when 
compared with their `field' equivalents, however the order of this 
difference is significantly smaller ($<$0.005~dex/kpc) than the 
systematic differences found between the RaDES and MUGS galaxies 
($\sim$0.05--0.2~dex/kpc). A comprehensive analysis of the (subtle) 
systematic differences between the field and loose group galaxies within 
RaDES is forthcoming (Few et al., in prep), but not pursued here, simply 
because this difference is negligible to the scope of the present 
analysis.}

We employed a strict kinematic decomposition of spheroid and disk stars 
for each of the 25 simulations\footnote{The kinematic decomposition 
employed for the MUGS galaxies differs from that used in the original 
\citet{Stinson10} analysis, in that $J_z$/$J_{circ}$ for each star 
was derived self-consistently taking into account the shape of the 
potential, rather than assuming spherical symmetry and using the 
enclosed mass at a given star particle's position.}, following the 
\citet{Abad03b} formalism. Additional (conservative) spatial cuts were 
employed to eliminate any satellite interlopers that might pass the 
initial kinematic decomposition. We define three age bins: young (stars 
born in the last 100~Myrs, to correspond roughly with B-stars), 
intermediate (stars formed 6$-$7~Gyr ago), and old (stars olders than 
10~Gyr).

\begin{figure}
\centering
\includegraphics[width=9cm]{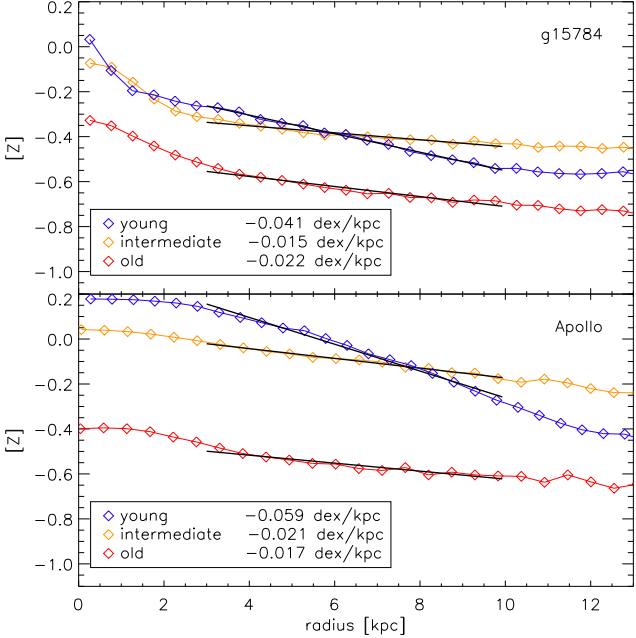}
\caption{Stellar radial [Z] gradients, for disk stars in three different 
stellar populations: young (blue) is defined as stars formed in the last 
100~Myrs, intermediate (yellow) is defined as stars formed 6 to 7~Gyr 
ago, and old (red) is defined as stars older than 10~Gyr. Fits to the 
disk are overdrawn in black; the length of the black line corresponds to 
the region of the disk used in the fitting (see text for details). For 
clarity, only two galaxies are shown, one from MUGS (g15784, upper 
panel) and one from RaDES (Apollo, lower panel).}
\label{star_radial}
\end{figure}

Observational studies of radial gradients typically show higher 
metallicities in the inner disk relative to the outer disk 
\citep[e.g.][]{Rup10}. As noted above, observations of external systems 
typically make use of gas-phase oxygen abundances, as measured from HII 
regions, but consistency exists between that tracer and others known to 
provide a ``snapshot'' of the gradient (e.g., planetary nebulae and 
short-lived main sequence B-stars).  Our gas-phase and young (B-star) 
gradients are identical in amplitude and gradient, and hence in what 
follows, we employ ``young stars'' (those formed in the previous 100~Myr 
period) to determine the abundance gradients.

The current RaDES sample only tracks global metallicity Z, but as oxygen 
consistently accounts for $\sim$50\% of Z, we use Z as a first-order 
proxy for oxygen, when making comparisions with 
observations.\footnote{We have recently completed the implementation of 
full chemical evolution, including SNeII, SNeIa, and AGB stars, within 
\textsc{ramses} - Few, Gibson \& Courty (2011, in prep).} The version of 
\textsc{gasoline} employed for these MUGS runs track both O and Fe (from 
SNeII and SNeIa), and assume Z$\equiv$O+Fe; as noted earlier, this 
latter assumption leads to an $\sim$0.2~dex underestimate of the global 
metallicity in the MUGS sample.  This does not impact upon our gradient 
analysis, but does serve to explain why the RaDES and MUGS galaxies are 
offset by $\sim$0.2~dex from one another in [Z] in the figures presented 
here.

Figure~\ref{star_radial} shows the mass-weighted radial gradients at 
$z$=0 in [Z] for one MUGS galaxy (g15784, top panel) and one RaDES 
galaxy (Apollo, lower panel). The radial gradients are calculated using 
linear fits over the noted disk regions (overdrawn in black). These are 
chosen to exclude the central region, avoiding any residual co-rotating 
bulge stars that escaped the kinematic decomposition. The outer edge of 
the disk is taken as the point at which the surface brightness profile
of the young stars (effectively, the cold gas)
deviates from an exponential. To 
ensure that an appropriate region is considered here, we have been 
conservative in choosing the ``disk region''. The gradient is robust to 
the choice of outer radius; reducing the choice of inner radius from 
5~kpc to 2~kpc has only a $\pm$0.007~dex/kpc impact on the inferred 
formal gradient - i.e., the differences in gradients between young, 
intermediate, and old populations are not significantly affected.
{Throughout this paper we use the \citet{Asp09} values for the 
solar metallicity.}

As one considers progressively older stellar populations (at the 
present-day), Figure~\ref{star_radial} shows that the measured radial 
metallicity gradient becomes progressively flatter. Such behaviour is 
not unexpected in cosmological simulations which include gas infall, 
radial flows, high velocity dispersion gas, kinematically hot disks, and 
dynamical mixing/radial migration which is more pronounced for older 
stars \citep[e.g.][]{San09,Ra11,Pilkington12}. 
The timescale of the mixing that flattens the gradients in the 
MUGS and RaDES simulations is shorter than the difference between 
intermediate and old populations of stars, as evidenced by radial 
gradients for the two populations, regardless of simulation suite, being 
quite similar.  The degree of flattening of the \it stellar \rm abundance
gradients is such that by the present day, \it within the simulations\rm, the
older stellar tracers show a flatter abundance
gradient than the younger tracers
(recall Fig~1, re-iterating results shown by \citet{San09}, 
\citet{Rah11}, and \citet{Pilkington12}).  This is counter 
to what is observed in the Milky Way when 
inferring gradients using younger planetary nebulae versus older planetary 
nebulae \citep[e.g.][]{Mac03}, but again, this is fully expected given the
degree of kinematic (stellar) heating within these cosmological 
simulations, and does \it not \rm impact on the use of gas-phase and
young-star probes of the gradients (both possess the expected steeper abundance
gradients at early-times).Indeed, future work in this area can, and should, 
make use of this powerful constraint on migration/heating: specifically, 
the fact that (empirically) older stellar probes today have a steeper 
abundance gradients than younger stellar probes, while extant, kinematically
hot, simulations, show the opposite trend.

For completeness, in Table~\ref{tabvals} we list the present-day 
mass-weighted stellar radial metallicity gradients (d[Z]/d$R$, in units 
of dex/kpc) for each of the 25 simulations employed here (column 8). The 
similarity of the gradients is readily apparent, save for the MUGS 
galaxy g24334, which was included in the sample despite its stellar 
fraction being dominated by accreted stars, rather than \it in situ\rm  
star formation (discussed further in \S~\ref{evo}). Its relatively small 
disk scalelength (1.0~kpc) also made fitting its gradient more 
challenging than the other MUGS disks.

Following \citet{San11}, we examined the effect of applying a different 
weighting scheme in determining the mean metallicities. When examining 
just the young stars or the gas, the weighting employed has no effect 
upon the inferred gradient.  However, when deriving a composite gradient 
making use of \emph{all} stars in the disk, the weighting can become 
important, as \citet{San11} suggested.  We explored the impact of using, 
for example, luminosity-weighting (and log-weighting), by deriving the 
absolute magnitude of each simulated star particle, making use of its 
age, metallicity, and initial mass function, alongside the 
\citet{Marigo08} 
isochrones.\footnote{\url{http://stev.oapd.inaf.it/cgi-bin/cmd_2.1}} As 
expected from the \citet{San11} analysis, the mean abundance shifted by 
$\sim$0.1~dex depending upon the weighting employed, but the inferred 
gradient was not affected.

\begin{figure}
\centering
\includegraphics[width=9cm]{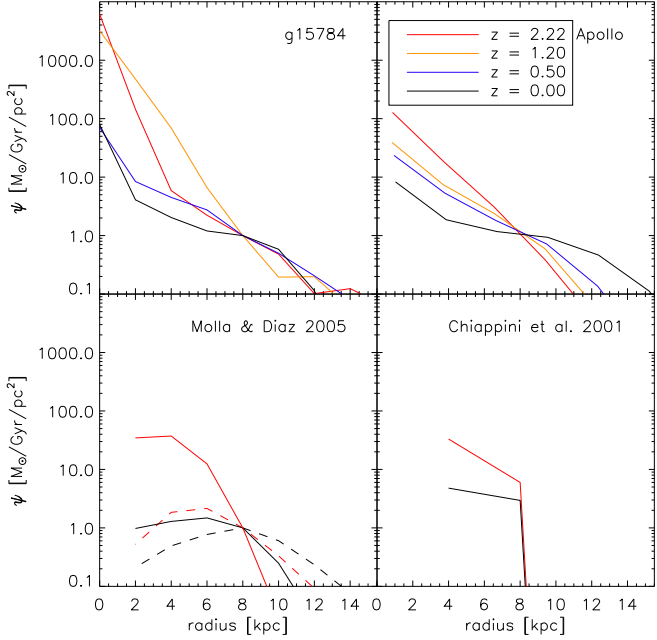}
\caption{Star formation rate per unit surface area as a function of 
radius for the MUGS galaxy g15784 (upper-left panel) and the RaDES 
galaxy Apollo (upper-right panel). We show the simulations at four 
different redshifts: $z$=0.0, 0.5, 1.2, and 2.2, as noted in the inset 
to the upper-right panel. 1~kpc annuli are used along with a height cut 
of $\pm$5~kpc above and below the disk. The mass of stars formed in the 
last 100~Myrs is calculated for each annulus out to a radius of 15~kpc. 
The curves have been normalised to 1~M$_{\odot}$/Gyr/pc$^2$ at 
galactocentric radius 8~kpc.  The bottom panels show the corresponding 
predicted behaviour of the \citet{Chiap01} (right) and \citet{Moll05} 
(left) models. Only redshifts 0.0 and 2.2 are shown, other redshifts are 
excluded as these models evolve smoothly from $z$=2.2 to $z$=0.0. Two of 
the \citet{Moll05} models are shown, one with high star formation 
efficiency (dashed lines) and one with low star formation efficiency 
(solid lines) .}
\label{sfrgrad}
\end{figure}

The abundance gradient of young stars (or equivalently, the ISM) is 
shaped by the time evolution of the radial star formation rate. To 
illustrate this we show the normalised star formation rate per unit 
surface area as a function of galactocentric radius in 
Figure~\ref{sfrgrad}. To match the chemical evolution models of 
\citet{Chiap01} for the Milky Way (with the understanding that our 
simulations are not constructed \it a priori \rm to be perfect replicas 
of the Milky Way), we normalise the star formation rate to have a value 
of 1~M$_{\odot}$/Gyr/pc$^2$ at a galactocentric radius of 
8~kpc.\footnote{ For context, the `normalised' and `pre-normalised'
star formation rate surface densities (at 8~kpc), for each of the
simulations, are not dissimilar; the latter lie 
in the range $\sim$1$-$2~M$_{\odot}$/Gyr/pc$^2$, save for the (known)
discrepant MUGS galaxy g24334 (which, pre-normalised, lies at
$\sim$0.2~M$_{\odot}$/Gyr/pc$^2$, reflective of the fact that its
stellar content is more dominated by its accreted component, rather
than in situ star formation.}

Each of the star formation rate profiles behave qualitatively like the 
classic inside-out chemical evolution models of \citet{Chiap01} and 
\cite{Moll05}, in the sense of decreasing outwards from the inner to 
outer disks.  An important systematic difference between these 
representative simulations is apparent though, at least at higher 
redshifts (1$<$$z$$<$2). Specifically, the gradient in the star 
formation rate per unit area is steeper at higher redshifts for the MUGS 
galaxies; it is not clear if this is symptomatic of a single difference 
between the MUGS and RaDES galaxies, or (more likely) a combination of 
factors including the star formation threshold, star formation 
efficiency, feedback schemes, and resolution of the respective 
simulations. Regardless, it is clear that \it star formation is more 
centrally-concentrated in the MUGS sample at early stages in the 
formation of the disk which unsurprisingly leads to steeper abundance 
gradients in the early disk \rm (a point to which we return shortly).

\subsection{Vertical Gradients}
\label{vertical}

\begin{figure}
\centering
\includegraphics[width=9cm]{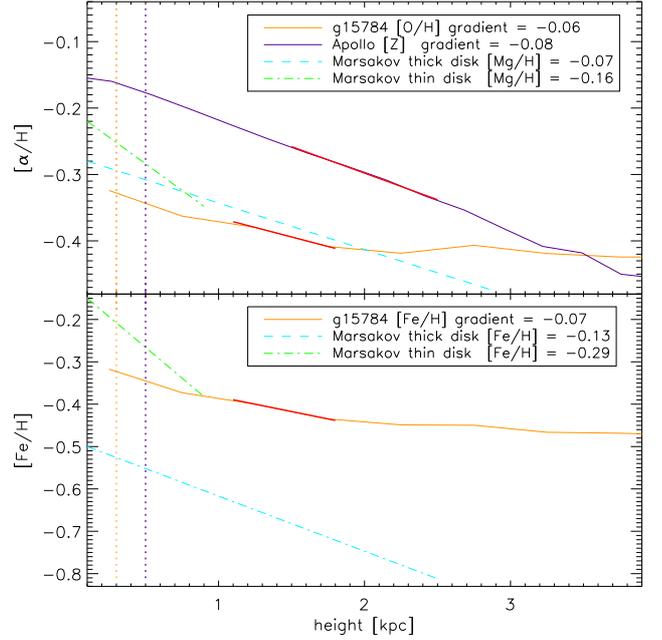}
\caption{The vertical gradients of disk stars in the simulations. The 
top panel shows the [Z] vertical gradient of Apollo (purple, grad = 
$-$0.08) with the [O/H] gradient of g15784 (orange, grad = $-$0.06) and 
observational data from \citet{Mar05,Mar06} of [Mg/H] gradients in the 
thin (blue, grad = $-$0.16) and thick (green, grad = $-$0.07) disk of 
the Milky Way. The lower panel shows the [Fe/H] gradients of the 
\citet{Mar05,Mar06} thin (grad = $-$0.29) and thick (grad =$-$0.13) disk 
data along with the g15784 (grad = $-$0.07) [Fe/H] gradient. Overplotted 
vertically are the softening length of the MUGS (orange) and the minimum 
grid size of the RaDES (purple) simulations. The bold red lines show the 
region used to calculate the gradient.}
\label{vertgrad}
\end{figure}

For completeness, as in Figs.~1 and 2, for g15784 (MUGS) and Apollo 
(RaDES), the mass-weighted vertical stellar abundance gradients in the 
simulations are presented in Figure~\ref{vertgrad}. A `solar 
neighbourhood' is defined for each simulation as being a 2~kpc annulus 
situated at a galactocentric radius of $\sim$2.5 disk scalelengths 
(column 5 of Table~\ref{tabvals}).  These radial scalelengths were 
derived from exponential fits to the stellar surface density profiles.

Classic work from, for example, \citet{Mar05,Mar06} and \citet{Sou08}, 
and soon-to-be-released work using SDSS-SEGUE and RAVE datasets, show 
that vertical metallicity profiles can provide extremely effective tools 
for separating the thin disk from the thick disk. With 
$\sim$300$-$500~pc softening/grid cells, we do not resolve the 
thin-thick disk transition. Figure~\ref{vertgrad}, shows the vertical 
gradient for the MUGS galaxy g15784 (orange) and the RaDES galaxy Apollo 
(purple), along with observational data for the Milky Way from 
\citet{Mar05} and \citet{Mar06}. The two vertical lines show the 
respective resolutions of the MUGS and RaDES simulations.

The vertical metallicity gradients (in their respective
`solar neighbourhoods') for the 25 simulations analysed here 
are listed in column 7 of Table~\ref{tabvals}. We find little
variation between the simulations in question, with the typical
vertical gradient lying in the $-$0.05$\pm$0.03~dex/kpc range.  Only
Eos, Castor, and Krios lie outside this range, possessing somewhat
steeper vertical abundance gradients.  These three undergo the 
most extended late-time period of `quiescent' evolution, as
commented upon in Few et~al., in prep.

At face value, the vertical gradients in [$\alpha$/H]\footnote{Here, 
total metallicity is used as a proxy for $\alpha$ in the RaDES suite, 
while oxygen is used for the MUGS and GCD+ suites; magnesium is used in 
the observational datasets described by \citet{Mar05,Mar06}.} and [Fe/H] 
inferred from the simulations are consistent with the observed values 
seen in the thick disk of the Milky Way ($\sim$$-$0.05 $-$ 
$\sim$$-$0.08~dex/kpc). The vertical gradients in the Milky Way's thin 
disk, though, are consistently much steeper (where many authors find the 
thin disk gradient to be between $\sim$$-$0.25 $-$ $\sim$$-$0.35~dex/kpc 
\citep[e.g.][]{Sou08,Mar06,Bart03,Chen03}) than the results we obtain 
from our simulations. Our spatial `resolutions' range from 
$\sim$300$-$500~pc, and the results appear compromised on vertical 
scales up to $\sim$2$-$3 resolution `elements' - i.e., any putative 
`thin' disk would be (not surprisingly) unresolved.  In a chemical 
sense, these disks are too `hot', in much the same way that their ISM 
and stellar populations are also kinematically hot 
\citep[e.g.][]{House11}.  

On this issue of `resolution', the
global star formation rates reported are comparatively well
converged as a function of resolution \citep[][\S5.2.4]{Stinson06}
The most notable change with
increasing resolution is the addition of higher redshift populations,
containing comparatively little mass, as earlier generations of halos
are resolved. This is at least partially a result of star formation
models largely being constrained to reproduce observed 
star formation rates.

The dependence of gradients on resolution though
is far less predictable. At our current
resolution we resolve sufficient substructure and disc dynamics to
capture the salient physical mechanisms involved in migration. However,
increasing resolution does resolve the physics behind migration
processes better, but it also 
makes the diffusion model more localized. Equally
importantly, it is not clear to what extent the numerous processes
involved in migration will interact with one another as resolution is
increased. Taking the alternative approach of lowering resolution makes
processes less likely to be captured (particularly substructure-induced
migration), so it is not clear that convergence happens in a simple
fashion. Ultimately, a definitive answer on the impact of resolution on
migration requires far higher resolution than we are currently able to
achieve and future work is required to address this issue. \rm

\section{Evolution of the Radial Gradients}
\label{evo}

While there exist a handful of studies of radial abundance gradients at 
high redshift \citep{Jones10, Cres10, Yuan11}, the difficulties in 
obtaining high resolution data for likely Milky Way-like
progenitors has meant that theoreticians have had 
very few constraints on their models; as noted earlier, inside-out 
galactic chemical evolution models can be constructed which recover the 
present-day gradients seen in the Milky Way, but they can take very 
different paths to get there. Some such models predict a steepening with 
time starting from initially inverted or flat gradients (e.g., 
\cite{Chiap01}), while others predict an initially negative gradient 
that flattens (e.g. \cite{Moll05}).

To make progress in this area, we now analyse the time evolution of the 
gradients within our 25 simulations, supplemented with two classical 
chemical evolution models, making fits radially at each timestep for 
which a clear disk could be identified.  As the disk is continually 
growing and evolving, we examined each timestep visually, identifying 
the outer `edge' using the cold gas and young stars as a demarcation 
point.  It should be noted here that the kinematic decomposition used to 
identify `disk stars' in \S~3.1 and \S~3.2 was not used for this 
component of our analysis.By working only with very young 
stars at 2$-$3 disk scalelengths, when fitting gradients at each 
timestep, kinematic decomposition of disk vs spheroid stars becomes 
unnecessary. Radial gradients were then derived by fitting typically 
from the outer edge of the disk to the inner part of the disk, where the 
inner point corresponds to the point at which the surface density 
profile deviates from an exponential.  Again, as we are only using the 
stars formed in the previous 100~Myrs (B-stars) at a given timestep, the 
relevant disk (rather than star-forming bulge) regime is not difficult 
to identify.

\begin{figure}
\centering
\includegraphics[width=9cm]{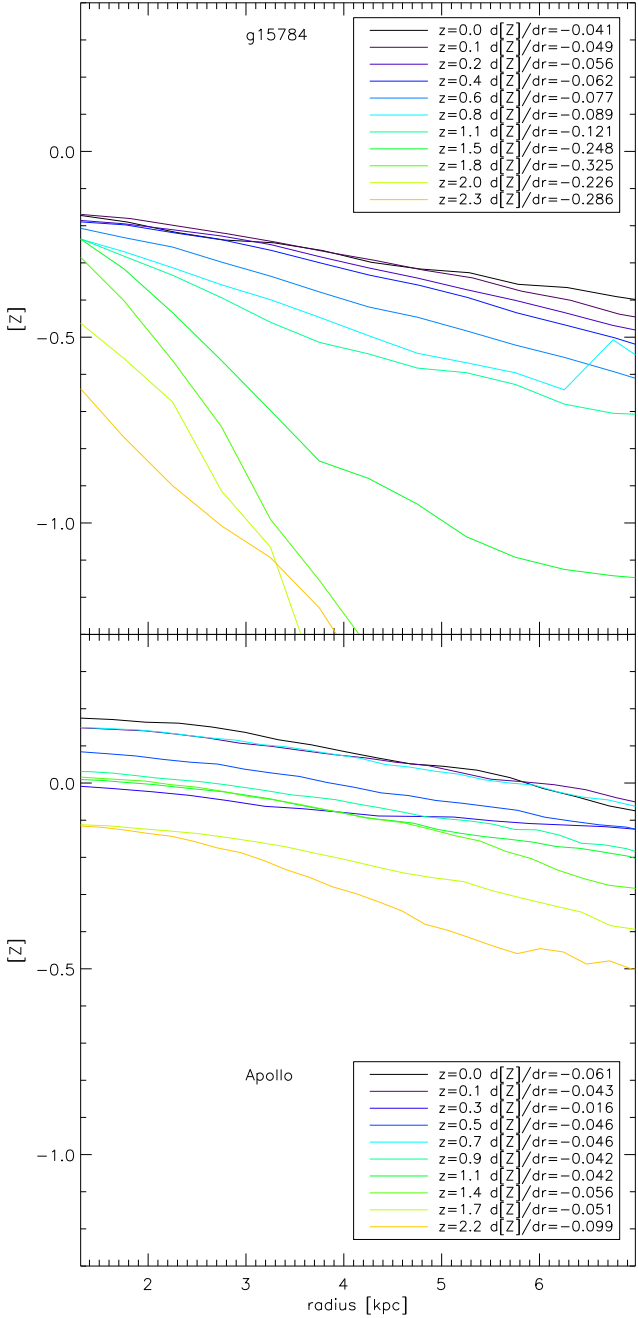}
\caption{The radial [Z] gradients of young stars in g15784 (top panel) 
and Apollo (bottom panel). The different colors correspond to different 
redshifts running from z$=$0 (black) to z$=$2.2 (orange), illustrating 
the time evolution of the abundance gradients in both simulations. Note 
the more dramatic flattening of the MUGS (g15784) relative to that of 
RaDES (Apollo). The fitted gradients were not done in an `automated' 
fashion; we examined each timestep's surface density, kinematic, and 
abundance profiles, to take into account the growth of the disk and 
identify the `cleanest' disk region within which to determine the 
gradient.}
\label{redshifts}
\end{figure}

In Figure~\ref{redshifts}, we show the time evolution of the radial 
gradient for our two `fiducial' simulations: MUGS (g15784, upper panel) 
and RaDES (Apollo, lower panel).  The gradients measured at each 
timestep are noted in the inset to each panel. Much steeper abundance 
gradients at high-redshift ($z$$>$1) are seen within the MUGS galaxy. 
Further, the offset in mean metallicity between the two, as already 
alluded to, can be traced to the manner in which chemistry was included 
in the version of \textsc{gasoline} employed (i.e., the assumption that 
Z$\equiv$O+Fe, which affects the mean metallicity, but not the 
gradient).

\begin{figure}
\centering
\includegraphics[width=9cm]{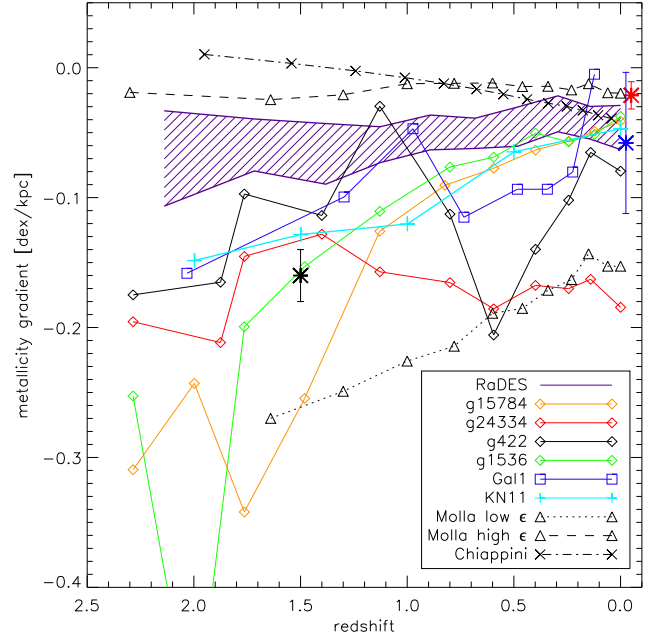}
\caption{The derived radial [Z] gradient as a function of redshift. 
Here, we have used 11 different redshifts and measured the radial 
gradient of the young stars (stars formed in the last 100~Myrs at each 
step) in the disk at that time. We examined the disks at each redshift, 
to determine the appropriate galactocentric radius over which to measure 
the gradients (see text for details). Four MUGS galaxies (g15784 (orange 
diamonds); g24334 (red diamonds); g422 (black diamonds); g1536 (green 
diamonds)) are shown, along with Gal1 (blue squares) from \citet{Rah11}, 
KN11 (cyan plus symbols) from \citet{Kob11},
and the 19 RaDES galaxies (denoted by the purple hatched area showing 
the region encapsulating 1$\sigma$ of the gradients measured at a given 
redshift). The two chemical evolution models are overlaid for 
completeness: Chiappini (black dot dashed crosses), and Moll\'{a} high 
efficiency (black dashed triangles) and low effiency (black dotted 
triangles). The black asterisk corresponds to the result from one lensed 
grand design spiral at $z$$\sim$1.5 \citep{Yuan11}, the blue asterisk to 
the typical gradient inferred in nearby spirals \citep{Zari94}, and the 
the red asterisk to the typical gradient seen in interacting disks 
\citep{Kew10}; these latter local points are offset slightly at $z$=0, 
for clarity.}
\label{age_grad}
\end{figure}

In Figure~\ref{age_grad}, we show the time evolution of the [Z] 
gradients for the 4 MUGS galaxies, the \textsc{gcd+} galaxy (Gal1), 
the \textsc{grape-SPH} galaxy (KN11),
and the 19 RaDES galaxies.  Importantly, we have also derived the time 
evolution of the predicted gradients for the chemical evolution models 
of \citet{Chiap01} and two of the Milky Way-like models of 
\citet{Moll05}; with the \citet{Moll05} data, the fits to determine the 
gradient at each timestep evolved as they did in the hydrodynamical 
simulations. As the disk grew, the fits were made at larger radii, to 
exclude the central region. From the earliest timestep to the latest the 
fitted region shifts $\sim$3~kpc in radius (reflecting the growth of the 
disk over the timescales under consideration). The \citet{Chiap01} data 
were fit over the radial range 4 to 8~kpc at each timestep, reflecting 
the fewer relevant annuli available over which to make the 
fit. \citet{Chiap01} fit their gradients to the same chemical 
evolution models over a broader radial range (4$-$14~kpc), but our 
interests here are restricted to the inner disks of these models, where 
the star formation density threshold is less important in shaping the 
metallicity gradient.

For the \citet{Moll05} models, we show a low-efficiency (28,8) and 
high-efficiency (28,2) example, (where model 28 corresponds to a 
circular velocity of $\sim$200~km/s and the efficiency factors 
correspond to the combined efficiency of molecular cloud formation and 
cloud-cloud collisions). The \citet{Chiap01} and, to a lesser extent, 
the high efficiency \citet{Moll05} models (at least since $z$$\sim$1) 
steepen with time.\footnote{The \citet{Chiap01} models have gradients 
which are mildly inverted at high-redshift ($\sim$$+$0.02~dex/kpc at 
redshift $z$$\sim$2); this works in the same direction as the inverted 
gradients observed by \citet{Cres10} at $z$$\sim$3, albeit the gradients 
claimed by the latter are significantly more inverted (i.e., 
$\sim$$+$0.1~dex/kpc) than encountered in any of the simulations 
or chemical evolution models. It is important to 
remember though that the AMAZE/LSD samples at $z$$\sim$3.3 are (a)
primarily Lyman-Break Galaxies with star formation rates 
($\sim$100$-$300~M$_\odot$/yr) well in 
excess of that expected for Milky Way-like progenitors, and are
not likely ideal progenitors against which to compare these 
simulations or chemical evolution models, and (b) in none of the 
current simulations are we able to unequivocally identify 
stable rotationally-supported disks, like those compiled by 
AMAZE/LSD.
We require targeted simulations with much 
higher resolution at high-redshift than we have access to here, and
tuned to be more representative of high-redshift Lyman-break
galaxies, before 
commenting further on this potentially interesting constraint.} 
Conversely, the RaDES sample (represented by the purple hatched region, 
which encompasses 1$\sigma$ of the gradient values at a given redshift) 
shows a mild flattening with time, more in keeping with full time 
evolution of the high efficiency \citet{Moll05} model.  The MUGS sample 
shows not only steeper gradients as a whole at $z$$>$1 (except for 
g24334, to which we return below), but also three of the four show the 
more significant degree of flattening alluded to in relation to 
Fig~\ref{redshifts}; this degree of flattening is more dramatic than 
that seen in any of the RaDES galaxies or the chemical evolution models 
(except for the low efficiency models of \cite{Moll05}).\footnote{It is 
worth noting that no obvious trend is seen when comparing the field and 
group galaxies in the RaDES sample. This is perhaps attributable to our 
selection criteria; by removing strongly interacting galaxies (at or 
near a pericentre passage), the sort of systematic differences seen in 
the work of \citet{Rupke10,Rup10,Perez11}, for example, would not be 
encountered here.}

Shown also in Fig~\ref{age_grad} are the typical gradients encountered 
in nearby isolated (\citet{Zari94}; blue asterisk) and interacting 
(\citet{Kew10}; red asterisk) disk galaxies (offset at $z$=0, for 
clarity, in Fig~\ref{age_grad}). The black asterisk at redshift 
$z$$\sim$1.5 corresponds to the recent determination of a steep 
metallicity gradient in a high-redshift grand design spiral by 
\citet{Yuan11}.  While intriguing, it is important to bear in mind that 
one should not necessarily make a causal link between these disparate 
data points; until a statistical sample of high-redshift gradients has 
been constructed, linking the \citet{Yuan11} point with those at 
low-redshift should be done with caution.

For this latter reason, we have also included one MUGS galaxy (g24334) 
in our analysis (red curve: Fig~\ref{age_grad}) that does not have a 
present-day gradient consistent with the typical late-type spiral.  We 
chose to include it, in order for the reader to see one example of a 
disk which possesses a steep gas-phase abundance gradient at 
high-redshift, comparable in slope to the \citet{Yuan11} 
observation, but one which does not evolve in time to resemble the 
shallower slopes seen in nature today. g24334 differs from the other 
MUGS galaxies, in the sense that the fraction of its stellar population 
born `in situ', as opposed to `accreted', is significantly lower.  
Further, its disk is less extended than the other Milky Way-analogs and 
its abundance gradient was derived at $\sim$0.5$\times$ disc 
scalelengths, where the gradient is more robust to interaction-induced 
flattening \citep[e.g.][]{Perez11}.

These differences are ultimately traced to the underlying treatment of 
star formation and feedback within the simulations; for example, the 
MUGS galaxies have a higher star formation threshold than the RaDES 
suite (1~cm$^{-3}$ vs 0.1~cm$^{-3}$).  As such, both the MUGS sample and 
the low efficiency models of \citet{Moll05} preferentially form stars in 
the inner disk where the densities are higher; the RaDES galaxies and 
the remaining chemical evolution models, with the lower threshold, have 
star formation occurring more uniformly throughout the early disk. 
Further, both MUGS and RaDES employ a standard blast-wave formalism for 
energy deposition into the ISM \citep{Stinson06}, but the latter imposes 
a minimum blast wave radius of 2 grid cells, which means that ejecta is 
in some sense more ``localised'' in the MUGS simulations (for the same 
SN energy, the RaDES blast waves are $\sim$2$-$3$\times$ larger); 
distributing energy (and metals) on larger radial scales can result in a 
more uniform (i.e., flattened) metallicity distribution. The trend of 
Gal1 lies somewhat between the extremes of MUGS and RaDES, which can be 
traced to the fact that Gal1 uses a lower star formation threshold 
density (0.1 cm$^{-3}$), and almost negligible feedback, resulting in 
more localised metal enrichment. KN11 also lies very close to the
MUGS fiducial (g15784) in terms of the temporal evolution of its
abundance gradient; both employ high SNe feedback efficiencies, albeit
on different spatial scales (a density-dependent blast wave radius in the
case of g15784 and a fixed 1~kpc radius in the case of KN11) and with
different star formation prescriptions (a 1~cm$^{-3}$ star formation
density threshold in the case of g15784 and an absence of a threshold
for KN11).  Note that although these hydrodynamical 
simulations experience different merger histories, the metallicity 
gradients are more affected by the recipe of sub-grid physics. This is 
highlighted by our large samples of simulations generated with different 
codes.

As detailed in \S~2.5, \citet{Chiap01} use a two infall model; at early 
times the infall of primordial gas is rapid and independent of 
galactocentric radius, while at later times, gas is assumed to fall 
preferentially on the outer regions of the disk, causing a steepening of 
the gradient with time. The radial dependence of this disk infall 
timescale is fairly gentle (linear with increasing radius); on the other 
hand, \citet{Moll05} calculate the overall infall rate as a function of 
the mass distribution and rotation of the galaxy, and assume a much 
stronger radial dependence for the infall timescale. Specifically, the 
inner disk's infall timescale is much more rapid than that of 
\citet{Chiap01}, while the outer disk's infall timescale is much longer.  
In combination, the gradient tends to flatten with time (particularly 
for their low efficiency models).

We find clear evidence of inside-out formation in the star formation 
profiles at different redshifts. Starting from an initially concentrated 
distribution, this flattens with time to the present-day, where star 
formation is more extended (and close to constant) over a large fraction 
of the disk (Fig~\ref{sfrgrad}). The radial dependence of star formation 
rate to infall rate sets the magnitude of the abundance gradient 
\citep{Chiap01}; a stronger radial dependence resulting in a steeper 
gradient. Such a configuration appears to come about naturally in the 
MUGS simulations, due in part to their higher star formation rate 
density threshold and perhaps the higher star formation efficiency and 
more localised chemical/energetic feedback. This contributes to the 
steeper gradients seen at early times in these simulations, relative to 
the other models. The RaDES galaxies behave more like the high 
efficiency model of \citet{Moll05}. It should be noted however that 
despite the significant differences seen in the early stages of these 
galaxies' evolution, the star formation distribution in the majority of 
these simulations is very similar at the present day.

\section{Summary}
\label{summ}

This work provides evidence in support of the \it imposed \rm inside-out 
disk growth paradigm adopted within chemical evolution models; this 
growth is a natural outcome of both Eulerian and Lagrangian 
hydrodynamical simulations of disk galaxy formation within a 
cosmological context. We have examined how this inside-out growth 
impacts on the magnitude and evolution of abundance gradients in these 
galaxies, using a suite of simulations and models which were calibrated 
to recover the present-day shallow gradients observed in late-type 
spirals. This is not meant to be a comprehensive, systematic, 
examination of sub-grid physics, in the vein of \citet{Wier11}, 
for example; instead, we have taken (in some sense) the `best' Milky 
Way-like simulations from several groups, using different codes, 
different initial conditions, and different assembly histories, and 
conducted a `blind' experiment on the outputs, to quantify \it how \rm 
the gradients evolved to the imposed boundary condition of a shallow 
present-day gradient.  Our findings include the following:

\begin{enumerate}

\item All galaxy models and simulations described in this work exhibit 
inside-out formation of the disk with varying degrees of 
centrally-concentrated star formation at early times 
(Figure~\ref{sfrgrad}). The evolving radial star formation rate 
dependence directly influences the resulting metallicity gradient; put 
another way, the signature of the star formation profile is embedded 
within the gradient of the young stars at each timestep. This signature 
though is diluted on the timescale of a few Gyrs. This is reflected in 
the differing gradients at the present-day between old and young stars 
(Figure~\ref{star_radial}); young stars at high-redshift within the MUGS 
sample (and observationally, it would appear, tentatively) form with a 
steep metallicity gradient, while those same stars today (now, old) have 
a fairly flat metallicity gradient (see Pilkington \& Gibson 2012).

\item Within the suite of 25 cosmological hydrodynamical simulations the 
derived vertical abundance gradients are comparable to those observed 
locally in the Milky Way's thick disk. The resolution is, however, not 
sufficient to discriminate between thin and thick disks.

\item The evolution of simulated metallicity gradients depends strongly 
on the choice of sub-grid physics employed and as such the magnitude and 
direction of its evolution depends critically upon the specific details 
of the recipes implemented. While it is difficult to disentangle the 
behaviour of the star formation profile {\it a priori\rm}, it is clear 
that simulated galaxies with more centrally-concentrated star formation 
have initially steeper abundance gradients. These are more consistent 
with the (albeit limited) observation of high redshift normal
Grand Design spiral galaxies \citep{Yuan11}. 

\item All the models and simulations tend to similar present-day 
abundance gradients, despite the diversity at earlier times, save for 
g24334 (which was chosen specifically in violation of the imposed 
shallow present-day gradient boundary condition, for illustrative 
purposes). In almost every case this requires the gradient to flatten 
with time, the exception being the chemical evolution model of 
\citet{Chiap01}. This model starts with an initially positive gradient 
that is independent of its halo phase. The gradient then inverts to 
become negative, with a gradient similar to other chemical evolution 
models.

\item The diversity of the evolution of metallicity gradient is for the 
first time highlighted by our large sample of both hydrodynamical 
simulations and chemical evolution models. Our results indicate that 
observations of the metallicity gradient for disk galaxies at different 
redshifts and that for the different age populations in the Galaxy are 
key to reveal the formation processes of disk galaxies and better 
constrain the sub-grid physics implemented with all the codes sampled.

\end{enumerate}

\noindent
Future work in this area will see us employ a finer temporal cadence, in 
order to better track the precise influence of merger events on the 
abundance gradients (both the magnitude of the effect and the timescale 
for re-establishing a stable abundance gradient). This study will also 
yield a deeper understanding of how the non-linear processes of star 
formation and feedback influence systematic differences between the 
various simulations presented here. We are near completion of a major 
upgrade to \textsc{ramses} which will allow us to re-simulate the RaDES 
suite with a broad spectrum of chemical elements, including those from 
SNeII, SNeIa, and AGB stars. With ongoing and future large scale 
spectroscopic surveys and missions such as RAVE, APOGEE, SEGUE, HERMES, 
LAMOST, and Gaia, providing detailed information on the phase and 
chemical space signatures of the Milky Way and beyond, such a 
chemodynamical exploration will be both timely and critical for 
understanding the origin and evolution of abundances in galaxies, and 
their link to the underlying physics of galaxy formation.

\begin{acknowledgements}
The authors would like to acknowledge Romain Teyssier, for ongoing 
access to \textsc{ramses}. We also acknowledge the superb guidance and 
support of Stephanie Courty, Mario Abadi, and Patricia Sanchez-Blazquez, 
throughout the prepation of this work. The exceptional critique 
provided by the referee is particularly appreciated. 
KP and CGF acknowledge the 
support of STFC through its PhD Studentship programme (ST/F007701/1); 
visitor support (LMD, DK, MM) from the STFC (ST/G003025/1) is similarly 
acknowledged. BKG, CBB, and DK, acknowledge the support of the UK's 
Science \& Technology Facilities Council (ST/F002432/1 \& ST/H00260X/1). 
LMD acknowledges support from the Agence Nationale de la Recherche 
(ANR-08-BLAN-0274-01). RJT acknowledges support from NSERC, CFI, the CRC 
program, and NSRIT.  BKG, KP, and CGF acknowledge the generous visitor 
support provided by Saint Mary's University. We thank the DEISA 
consortium, co-funded through EU FP6 project RI-031513 and the FP7 
project RI-222919, for support within the DEISA Extreme Computing 
Initiative, the UK's National Cosmology Supercomputer (COSMOS), the 
University of Central Lancashire's High Performance Computing Facility, 
CfCA/NAOJ, JSS/JAXA, and the Shared Hierarchical Academic Research 
Computing Network (SHARCNET).
\end{acknowledgements}

\bibliographystyle{aa} 
\bibliography{Gradients}

\end{document}